\documentclass[onecolumn,aps,pra,superscriptaddress,preprint]{revtex4-1}
\usepackage{amsmath,latexsym}
\usepackage{xcolor}
\usepackage[%
colorlinks=true,
urlcolor=blue,
linkcolor=blue,
citecolor=blue
]{hyperref}
\usepackage{etoolbox}

\makeatletter
\let\cat@comma@active\@empty
\makeatother

\usepackage{graphicx}
\usepackage{booktabs}
\AtBeginDocument{
\heavyrulewidth=.08em
\lightrulewidth=.05em
\cmidrulewidth=.03em
\belowrulesep=.65ex
\belowbottomsep=0pt
\aboverulesep=.4ex
\abovetopsep=0pt
\cmidrulesep=\doublerulesep
\cmidrulekern=.5em
\defaultaddspace=.5em
}

\usepackage{threeparttable,array,booktabs,calc}
\usepackage{graphicx}
\usepackage{dcolumn}
\usepackage{bm}

\usepackage[version=4]{mhchem}

\usepackage{hyperref}

\usepackage{epstopdf} 

\usepackage[official]{eurosym}

\begin{document}

\title[Optical quenching of singlet-He]{Optical quenching of metastable helium atoms using excitation to the 4P state}

\author{Jiwen Guan}
\thanks{Both authors contributed equally to this work.}
\affiliation{Institute of Physics, University of Freiburg, Hermann-Herder-Str. 3, 79104 Freiburg, Germany}
\author{Vivien Behrendt}
\thanks{Both authors contributed equally to this work.}
\affiliation{Institute of Physics, University of Freiburg, Hermann-Herder-Str. 3, 79104 Freiburg, Germany}
\author{Pinrui Shen}
\affiliation{Department of Physics and Astronomy, University of British Columbia, Vancouver, BC, V6T 1Z1, Canada}
\author{Simon Hofs\"ass}
\altaffiliation{Current address: Fritz-Haber-Institut der Max-Planck-Gesellschaft, Faradayweg 4-6, 14195 Berlin, Germany}
\affiliation{Institute of Physics, University of Freiburg, Hermann-Herder-Str. 3, 79104 Freiburg, Germany}
\author{Thilina Muthu-Arachchige}
\affiliation{Institute of Physics, University of Freiburg, Hermann-Herder-Str. 3, 79104 Freiburg, Germany}
\author{Jonas Grzesiak}
\affiliation{Institute of Physics, University of Freiburg, Hermann-Herder-Str. 3, 79104 Freiburg, Germany}
\author{Frank Stienkemeier}
\affiliation{Institute of Physics, University of Freiburg, Hermann-Herder-Str. 3, 79104 Freiburg, Germany}
\author{Katrin Dulitz}
\email{katrin.dulitz@physik.uni-freiburg.de.}
\affiliation{Institute of Physics, University of Freiburg, Hermann-Herder-Str. 3, 79104 Freiburg, Germany}

\date{\today}

\begin{abstract}
Discharge and electron-impact excitation lead to the production of metastable helium atoms in two metastable states, 2$^1$S$_0$ and 2$^3$S$_1$. However, many applications require pure beams of one of these species or at least a detailed knowledge of the relative state populations. In this paper, we present the characterization of an original experimental scheme for the optical depletion of He(2$^1$S$_0$) in a supersonic beam which is based on the optical excitation of the 4$^1$P$_1 \leftarrow 2^1$S$_0$ transition at 397 nm using a diode laser. From our experimental results and from a comparison with numerical calculations, we infer a near unit depletion efficiency at all beam velocities under study (1070 m/s $\leq v \leq$ 1750 m/s). Since the technique provides a direct means to determine the singlet-to-triplet ratio in a pulsed supersonic helium beam, our results show that the intrabeam singlet-to-triplet ratio is different at the trailing edges of the gas pulse. 
\end{abstract}


\maketitle

\section{Introduction}
Atoms and molecules in electronically excited states are considered as ``metastable", if transitions to lower-lying energy states are forbidden by electric-dipole selection rules. Because of their long natural lifetimes ($\geq 10^{-5}$ s) and their high internal energy ($\geq$ 10 eV), the metastable states of the noble gases play an important role in a variety of environments including planetary atmospheres, flames and plasmas \cite{Makabe1992,Falcinelli2015}.

Over the years, there has been continuing interest in the use of metastable species for a variety of applications \cite{Gay1996}. For example, metastable noble-gas species serve as sensitive probes in surface analysis techniques, such as metastable atom electron spectroscopy (MAES) and metastable de-excitation spectroscopy (MDS) \cite{Harada1997, Onellion1984}. Metastable noble gases have laser-accessible, electric-dipole-allowed transitions to higher-lying states which also make them particularly suitable for use in laser cooling \cite{Vassen2012}. Applications of such laser-cooled species include the production of ultracold, quantum degenerate gases, precision spectroscopy and atomic interferometry \cite{Vassen2012} (and references therein), \cite{Vassen2016}. Traditionally, metastable helium atoms have also been popular test systems for new laser cooling schemes, e.g., for sub-Doppler cooling by velocity-selective coherent population trapping \cite{Aspect1988, Hack2000}, for white-light cooling \cite{Rasel1999} and for bichromatic cooling \cite{Cashen2001,Partlow2004,Corder2015}. The velocity of metastable helium atoms in a supersonic beam has also been successfully manipulated using the Zeeman deceleration technique which does not rely on laser cooling \cite{Dulitz2015a}.

Metastable noble gases have been subject to many fundamental studies of autoionizing collisions at thermal and at ultracold temperatures \cite{Vassen2012, Siska1993}. For example, recent studies of reactive collisions with metastable noble gases in merged supersonic beams have enabled the observation of quantum resonances and stereodynamic effects \cite{Henson2012, Gordon2017}. In fact, the quantum degeneracy of metastable noble gases could only be achieved by electron-spin polarizing the atoms which suppresses ionizing collisions \cite{Fedichev1996, Herschbach2000}.

Metastable noble gases can be produced in a variety of ways, including electron-beam bombardment, discharge excitation, charge transfer, optical pumping, and thermal excitation \cite{Gay1996}. Among the sources, the electron-beam bombardment and discharge sources are most commonly used. Since the electron excitation energies are typically well above threshold in order to maximize the metastable production rate \cite{Rundel1974a,Dunning1975}, two metastable states are populated simultaneously, the 2$^3$S$_1$ and 2$^1$S$_0$ states in the case of helium and the $^3$P$_0$ and $^3$P$_2$ states in the case of the heavier noble gases. However, many experiments require either pure beams of a single metastable species or a precise knowledge of the relative state populations. Beam purification can be achieved in several ways, e.g., by beam deflection in a magnetic field \cite{Weiser1987}, coherent momentum transfer \cite{Theuer1998} or by transverse optical deflection \cite{Aspect1990}. State purification is also achieved by optical pumping (often denoted as ``quenching"). In this scheme, the population of one metastable state is transferred to the electronic ground state via optical excitation to a higher-lying electronic state, whose decay to the electronic ground state is strongly allowed for electric-dipole radiation. There exist efficient optical pumping schemes for atomic beams of metastable helium, neon, argon and krypton \cite{Fry1969, Hotop1969a, Hotop1981, Harada1987,Dunning1975, Gaily1980, Verheijen1986, Brand1992, Kau1998, Thiel2004}. The de-excitation of metastable helium atoms in the 2$^1$S$_0$ state has thus far only been attained in a supersonic beam by illuminating the atomic beam with 2058 nm light from a helium discharge lamp, which is resonant with the 2$^1$P$_1 \leftarrow 2^1$S$_0$ transition \cite{Fry1969, Kato2012, Hotop1969a, Hotop1981, Harada1987}. The use of a discharge lamp requires a complex, bulky setup with a coil-shaped, water-cooled lamp that encloses the supersonic beam inside the vacuum chamber. Apart from its complexity, such a setup can strongly perturb the supersonic flow and thus degrade the performance of the beam source. If the water-cooling of the lamp is not sufficient, such a lamp may easily heat up and thus prevent its continuous operation.

In this work, we demonstrate an original and very efficient approach to the optical depletion of He(2$^1$S$_0$) in a supersonic beam using optical excitation via the 4$^1$P$_1 \leftarrow 2^1$S$_0$ transition at a vacuum wavelength of 396.58509 nm \cite{Martin1960}. This scheme is based on simple and inexpensive diode laser technology, which makes it suitable for many different experimental applications. We provide a detailed experimental characterization of the technique, including an examination of the quenching efficiency as a function of laser power, detuning from resonance and helium beam velocity. We also give an account of the numerical calculations used for the determination of the depletion efficiency.

\section{Experiments}
\begin{figure}[h!]
	\includegraphics[width=\linewidth]{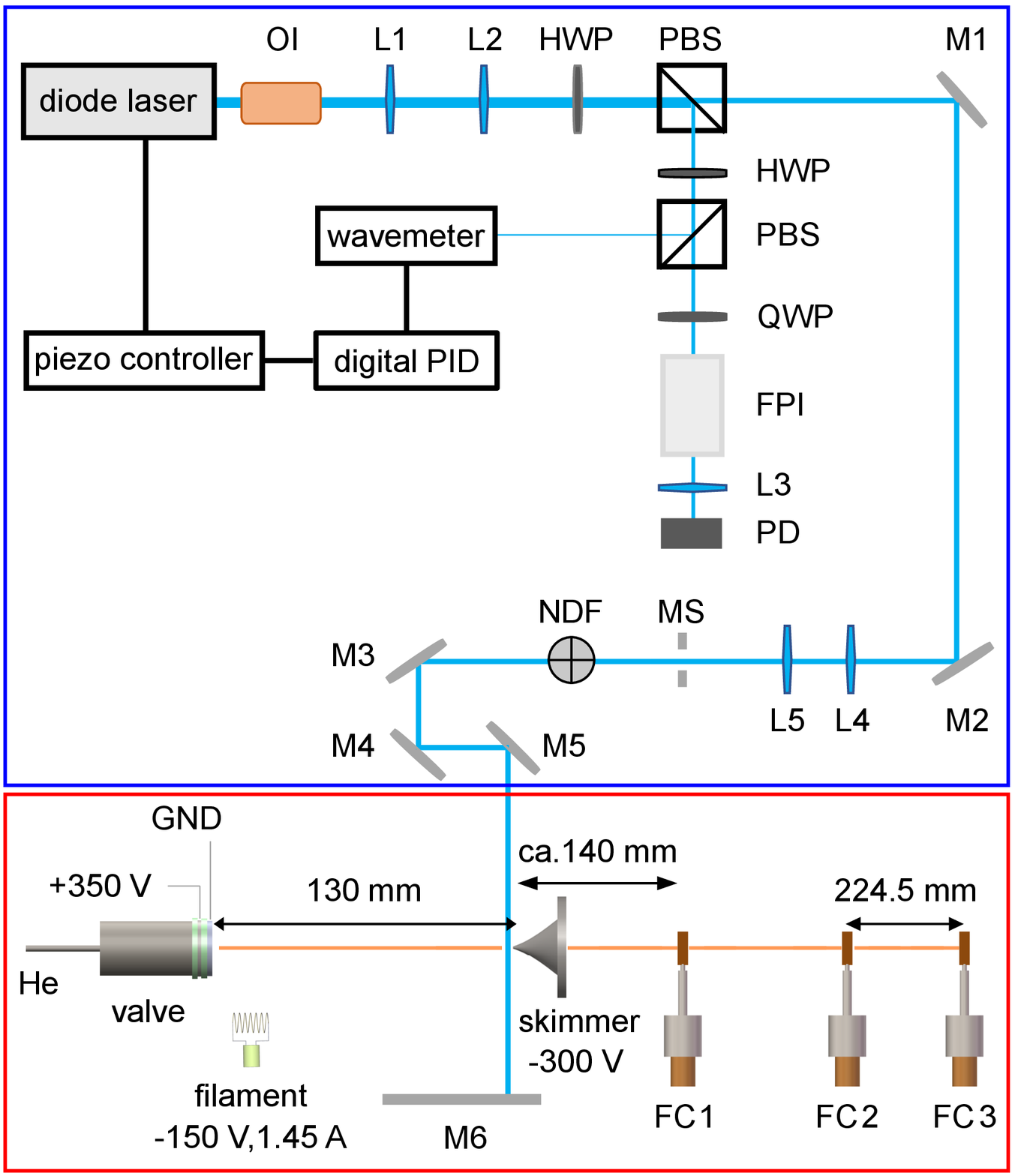}
	\caption{Schematic drawing of the experimental setup which consists of the optical system (blue box) and a vacuum apparatus (red box). Abbreviations: OI = 30 dB optical isolator, L1--L5 = lenses, M1--M6 = mirrors, HWP = half-wave plate, QWP = quarter-wave plate, PBS = polarizing beamsplitter, FPI = Fabry-P\'{e}rot interferometer, PD = photodiode, MS = mechanical shutter, NDF = neutral-density filter, PID = proportional-integral-derivative controller, GND = chassis ground, FC = Faraday cup.}
	\label{fig:setup}
\end{figure}

A schematic drawing of the experimental setup, which consists of the optical system and a vacuum apparatus, is shown in Fig. \ref{fig:setup}. In the following, both parts will be described in detail.

\subsection{Vacuum Apparatus}
Experiments are carried out in a set of three vacuum chambers. Inside the first chamber, a pulsed supersonic beam of metastable helium atoms is produced by a He gas expansion into vacuum and subsequent discharge excitation. The supersonic beam is generated by a high-intensity, short-pulse solenoid valve (CRUCS, $d = 100$ $\mu$m orifice diameter, $40^{\circ}$ cone, copper body) whose characteristics are described in Ref. \cite{Grzesiak2018}. The temperature of the valve is controlled by a cryocooler (CTI, 350CP) whose temperature is typically regulated to a set value with accuracy to within 0.1 K using proportional-integral-derivative (PID)-controlled resistive heating (LakeShore Model 325). The discharge unit used for the excitation of ground-state helium atoms into the metastable 2$^3$S$_1$ and 2$^1$S$_0$ states is described in a previous publication \cite{Grzesiak2019} and is thus not detailed here. A pulse duration of $\leq 30\,\mu$s is inferred from the sudden decrease of the plate voltage upon discharge excitation. At each valve temperature, the He stagnation pressure and the settings at the valve driver were adjusted to maximize the metastable helium signal intensity and to avoid bouncing of the valve plunger. For the measurements on optical quenching described here, valve stagnation pressures between 10--30 bar were used.

Laser excitation is done at a distance of $z_0 \approx 130$ mm from the valve exit, in close proximity to the skimmer ($b = 1$ mm diameter, see Fig. \ref{fig:setup}). This arrangement avoids a repopulation of the 2$^1$S$_0$ state by the discharge and it ensures a good spatial overlap between the laser beam and the He beam, with the tip of the skimmer serving as a handy tool for the alignment of the laser beam with respect to the He beam. After laser excitation, the beam of metastable helium atoms is monitored on a copper plate, which serves as a Faraday-cup-type detector \cite{Hotop1996} (denoted as FC 1 in Fig. \ref{fig:setup}), inside the second chamber. The signal is amplified using a transimpedance amplifier (Femto, DLPCA-200, $10^6$ V/A gain, 500 kHz). The detection efficiency of such a detector is not well known ($\approx 50$ \% \cite{Siska1993}). However, only relative signal intensities with and without laser irradiation are measured so that the absolute detection efficiency is not important for this work. To avoid the detection of ions and atoms in Rydberg states produced during the discharge process, the skimmer is biased to a voltage of -300 V relative to chassis ground.

Another set of Faraday-cup-type detectors (FC 2 and FC 3 in Fig. \ref{fig:setup}) is placed inside a third chamber. Since their relative distance is accurately known (224.5$\pm$0.5 mm), the longitudinal velocity of the supersonic beam of metastable helium atoms can be inferred from the difference $\Delta t$ of the time-of-flight signal intensities at the two detectors \cite{Grzesiak2018}.

During the experimental runs, the three vacuum chambers are held at pressures of approximately \mbox{$2 \cdot 10^{-6}$ mbar}, \mbox{$2 \cdot 10^{-7}$ mbar} and \mbox{$3 \cdot 10^{-10}$ mbar}, respectively.

\subsection{Optical System}
Laser radiation at 397 nm, used to drive the 4$^1$P$_1 \leftarrow 2^1$S$_0$ transition, is generated inside a standard, home-built external cavity diode laser system (NDU4316 laser diode by Nichia). To allow for a straightforward frequency tuning of the diode laser, the laser frequency is locked to the output of a wavelength meter (High Finesse, WS7, 60 MHz absolute accuracy, 2 MHz relative accuracy) using digital PID control. For this, the difference between the measured wavelength and the user-set wavelength is converted into an error signal and a corresponding analog voltage using MATLAB code and a digital-to-analog converter (DAC). This voltage is used to control the piezo element which adjusts the laser grating, and hence, the laser wavelength. The mean sampling rate and the stability of the laser lock are $\approx 100$ Hz and 2 MHz, respectively. \footnote{In this estimate, frequency drifts caused by thermal fluctuations inside the wavelength meter itself are disregarded.} The absolute frequency of the wavelength meter was initially obtained by comparing the wavenumbers of the two hyperfine-structure components and the crossover resonance in the $^7$Li D$_2$ line \cite{Sansonetti1995}, obtained by Doppler-free frequency-modulation spectroscopy, with the output of the wavelength meter. During the optical depletion experiments, the resonance frequency is obtained from the measured line profile of the 4$^1$P$_1 \leftarrow 2^1$S$_0$ transition.

A Fabry-P\'{e}rot interferometer (FPI, finesse $\mathcal{F} \approx 48$) and a photodiode (PD) are used to monitor the single-mode operation of the laser. A lens is attached to the entrance of the FPI unit to achieve the mode-matching for the cavity (not shown in Fig. \ref{fig:setup}). In addition to that, a quarter-wave plate (QWP) is inserted in front of the interferometer to prevent back-reflections into the laser diode. 

The remaining laser radiation is overlapped with the He beam at right angles to minimize Doppler broadening of the 4$^1$P$_1 \leftarrow 2^1$S$_0$ transition originating from the longitudinal velocity distribution of the supersonic beam. To increase the interaction time with the sample, the laser beam is retro-reflected through the He beam using a high-reflectivity mirror (M6 in Fig. \ref{fig:setup}). The mirror is located inside the vacuum chamber to avoid transmission losses through the optical windows of the vacuum chamber. The incoming light is sent at a very small vertical angle with respect to the retro-reflected beam to avoid back reflections into the diode laser.

To minimize influences by shot-to-shot fluctuations of the helium beam source, the laser beam was toggled between open and closed in between two subsequent shots of the pulsed valve using a fast mechanical shutter (SRS, SR475). A servo-motor-controlled neutral-density filter wheel allowed for a fast and accurate setting of the laser power (between 0--50 mW) admitted into the chamber. The laser power was measured on a power meter (Coherent, OP-2 VIS) outside the vacuum chamber. Transmission losses at the entrance window into the vacuum chamber (9 \%) were taken into account in the analysis. The semi-axes of the elliptical laser beam parallel and perpendicular to the He beam propagation axis were determined at the position of the interaction region as $a = 0.54$ mm  and $b = 0.23$ mm  (intensity full width at half maximum (FWHM)), respectively, using a beam profiler (LaserCam-HR, Coherent).

\section{Numerical Calculations}
\begin{figure}[h!]
	\includegraphics[width=\linewidth]{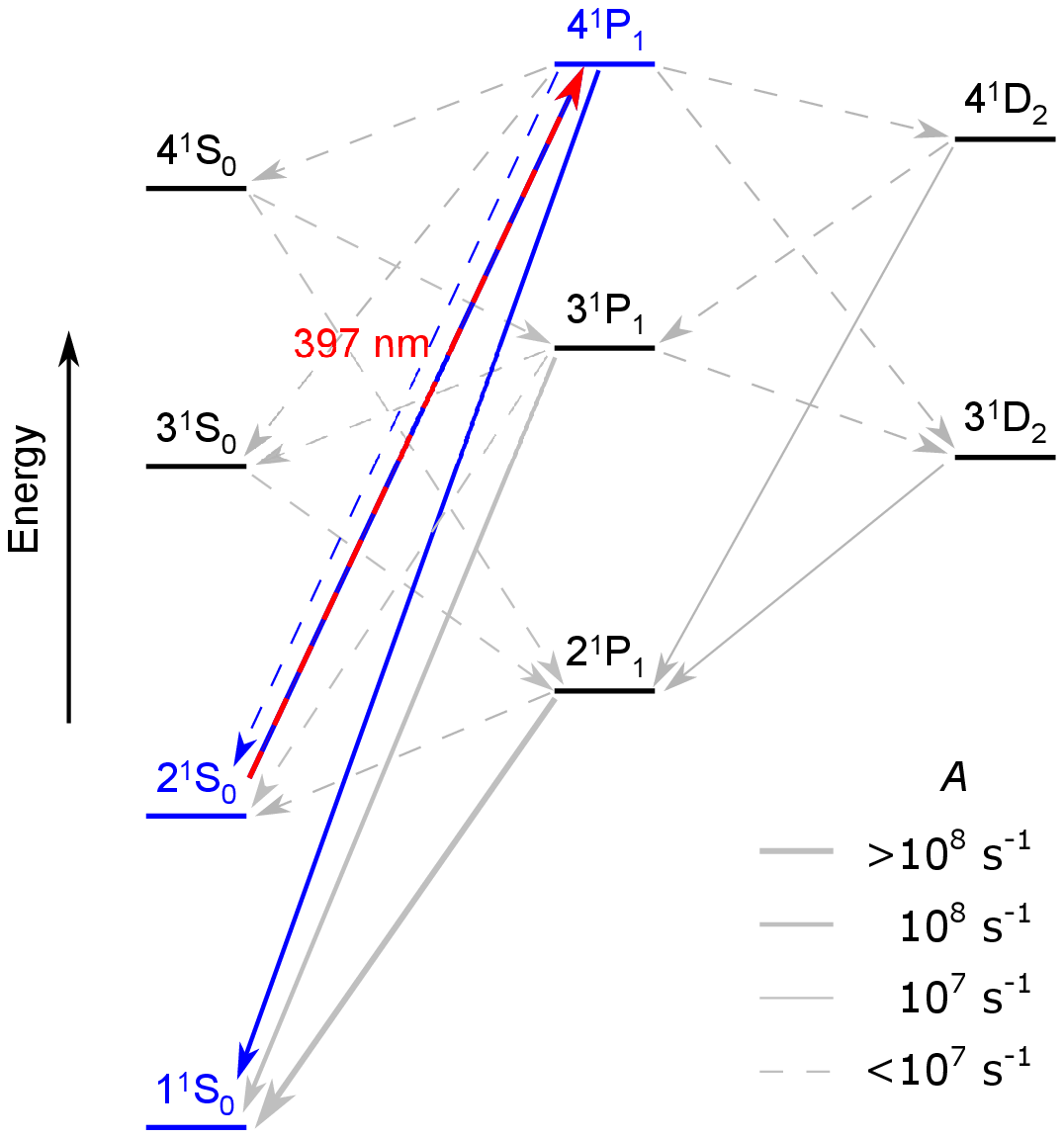}
	\caption{Schematic representation of the He energy levels and the transitions involved in the optical depletion scheme. The 4$^1$P$_1 \leftarrow 2^1$S$_0$ transition at 397 nm is shown as an arrow in red and blue color. All electric-dipole-allowed decay routes from the 4$^1$P$_1$ state are also indicated as arrows. The linestyle of these arrows provides a rough estimate of the transition probability, expressed as Einstein A coefficients. All transitions taken into account in the three-level model are shown in blue color.}
	\label{fig:ratemodel}
\end{figure}

To quantify the depletion efficiency, we have numerically solved the rate equations for the system using Mathematica and Matlab codes following standard procedures. Even though excitation to the 4$^1$P$_1$ state involves decay routes via eight electronic states, as shown in Fig. \ref{fig:ratemodel}, the process can be accurately described by a three-level model (marked in blue color in Fig. \ref{fig:ratemodel}). This can be rationalized by considering that the Einstein A coefficient for the 4$^1$P$_1 \rightarrow 2^1$S$_0$ transition is a factor of 35 higher than for all other transitions from the 4$^1$P$_1$ state. A direct comparison also shows that the population difference obtained from the solution to the nine-level model and to the three-level model is negligible (less than 1 \%) under all conditions studied here.

The rate equations for interaction with non-polarized light can be written as follows \cite{Metcalf1999, Steck2001, Budker2004}:
\begin{subequations}\label{eq:rateeq}
\begin{align}
\frac{\mathrm{d}N_{i}}{\mathrm{d}t} &= -\Gamma_{ie}N_{i}(t) +A_{ei}N_{e}(t) +\Gamma_{ei}N_{e}(t)\\
\frac{\mathrm{d}N_{e}}{\mathrm{d}t} &= +\Gamma_{ie}N_{i}(t) -A_{ei}N_{e}(t) -\Gamma_{ei}N_{e}(t) -A_{ef}N_{e}(t)\\
\frac{\mathrm{d}N_{f}}{\mathrm{d}t} &= +A_{ef}N_{e}(t),
\end{align}
\end{subequations}
where $N_{i}$, $N_{e}$ and $N_{f}$ denote the populations of the 2$^1$S$_0$ state, the 4$^1$P$_1$ state and the 1$^1$S$_0$ state, respectively. At $t = 0$, all the population is assumed to be in state $i$, i.e., $N_{i} = 1$. The Einstein A coefficients were taken from the NIST Atomic Spectra Database \cite{NIST_ASD}. During laser irradiation, the pump rate $\Gamma_{ei} = \Gamma_{ie}$ is given by
\begin{equation}\label{eq:gamma}
\Gamma_{ei} = \frac{3c^2}{2h\pi\nu_0^3}\frac{P_\mathrm{l}}{A_\mathrm{l}}\frac{1}{1+(4\pi\delta\nu/\left(A_{ei}+A_{ef}\right))^2}.
\end{equation}
It depends on the incident laser power $P_\mathrm{l}$, the transition frequency $\nu_0$, the frequency detuning from resonance $\delta\nu$ and the interaction area $A_\mathrm{l}$. To simplify the calculation, an elliptically shaped light beam with a homogeneous intensity distribution is assumed, so that $A_\mathrm{l} = \pi a b$, where $a$ and $b$ are the semi-axes of the laser beam at FWHM (see above). A solution of the rate equations in the presence of the light field, and taking into account the population relaxation into a new equilibrium after laser irradiation (where $\Gamma_{ei} = 0$), results in a Lorentzian line profile which accounts for the natural linewidth and for power broadening.

The line profile, obtained from the procedure above, is then convoluted with a Gaussian distribution of standard deviation $\sigma = \sqrt{\sigma_{\mathrm{D}} + \sigma_{\mathrm{l}}}$ to take into account Doppler broadening ($\sigma_{\mathrm{D}}$) and the laser linewidth ($\sigma_{\mathrm{l}}$) as additional line-broadening mechanisms. In our experiment, the contribution from Doppler broadening is due to the transverse velocity component of the supersonic beam. The Doppler linewidth is linearly dependent on the helium beam velocity $v$. The Doppler width at FWHM is assumed as \cite{Hollenstein2003}
\begin{equation}\label{eq:Doppler}
\Delta v_{\mathrm{D}} = 2\sqrt{2\ln(2)}\nu_0\frac{ v\sin{(\beta})}{c},
\end{equation}
where $c$ is the speed of light, and the opening angle
\begin{equation}\label{eq:openingangle}
\beta = \arctan{\left( \frac{1}{2} \frac{d+b}{z_0}\right) }
\end{equation}
depends on the orifice diameter $d$, the skimmer diameter $b$ and the distance between nozzle and skimmer $z_0$. A Doppler shift of the spectral line is not taken into account owing to the close-to-perpendicular geometry between the supersonic beam and the laser beam. Contributions by other line-broadening mechanisms are negliglibly small and thus not taken into account.

\section{Results and Discussion}
\begin{figure}[h!]
	\includegraphics[width=\linewidth]{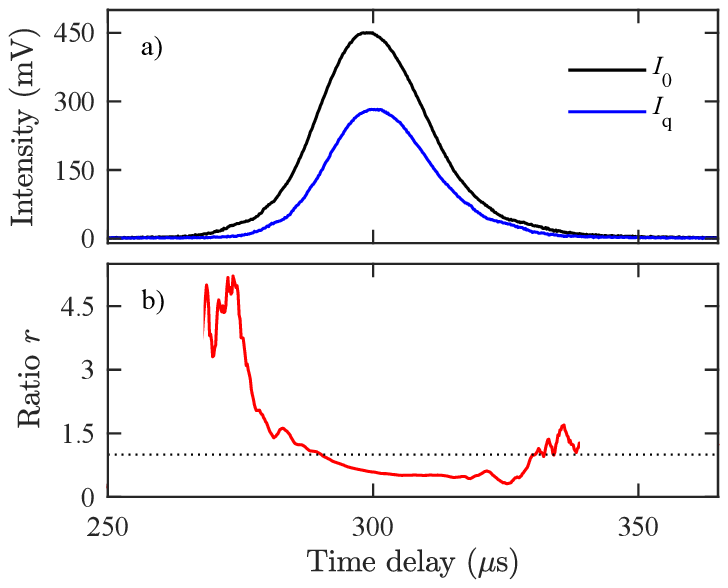}
	\caption{(a) Time-of-flight traces of metastable helium atoms ($v$ = 1070 m/s) measured at FC 1 in the presence (blue line) and in the absence (black line) of 38 mW of laser light resonant with the 4$^1$P$_1 \leftarrow 2^1$S$_0$ transition. The time delay is given with respect to the valve trigger pulse. (b) Helium singlet-to-triplet ratio (obtained using Eq. \ref{eq:ratio}) as a function of beam time-of-flight. Ratios obtained from low signal intensities are not shown for clarity. The dotted line at $r = 1$ is for visibility only.}
	\label{fig:TOF}
\end{figure}

\begin{figure}[h!]
	\includegraphics[width=\linewidth]{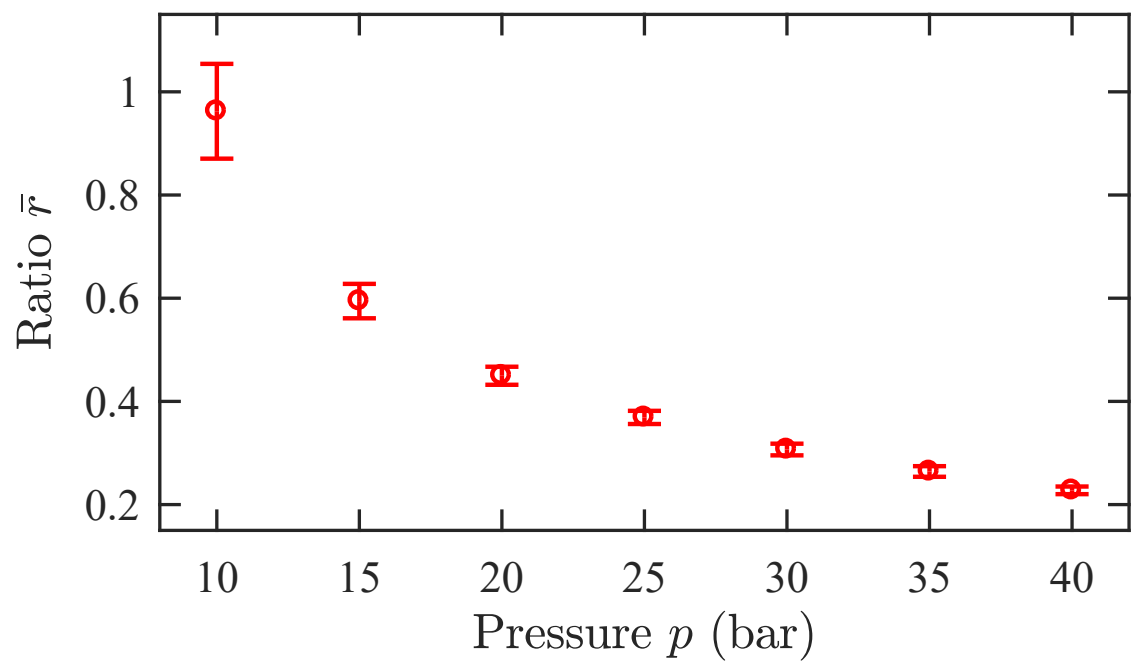}
	\caption{Average helium singlet-to-triplet ratio as a function of valve stagnation pressure at $v$ = 1070 m/s and at a laser power of 30 mW.}
	\label{fig:stratio}
\end{figure}

Fig. \ref{fig:TOF} (a) shows time-of-flight (TOF) traces of metastable helium atoms in the presence (blue line, signal intensity $I_{\mathrm{q}}$) and in the absence (black line, signal intensity $I_{0}$) of laser radiation resonant with the 4$^1$P$_1 \leftarrow 2^1$S$_0$ transition. Since Faraday-cup detection measures both He(2$^1$S$_0$) and He(2$^3$S$_1$) signal contributions, the signal intensity does not go to zero in the presence of laser light, even though the laser power of 38 mW used here leads to a full depletion of the 2$^1$S$_0$ state population at $v$ = 1070 m/s (see discussion below). When measuring under conditions at which the population in the $2^1$S$_0$ state is fully depleted, the remaining signal intensity originates from the He(2$^3$S$_1$) state only. Therefore, the signal ratio $I_{\mathrm{q}}/I_{0}$ can be directly related to the helium singlet-to-triplet ratio in the supersonic beam
\begin{equation}\label{eq:ratio}
r = \frac{I_{0}-I_{\mathrm{q}}}{I_{\mathrm{q}}}
\end{equation}
if the same detection efficiency is assumed for both metastable states.
As can be seen from Fig. \ref{fig:TOF} (b), the singlet-to-triplet ratio is much higher at the rising edge of the helium gas pulse than at other times. This effect may be related to a more efficient \textit{collisional} quenching of He(2$^1$S$_0$) by thermal electrons in the central, higher-density part of the beam. These thermal electrons are produced during the discharge process and may lead to a de-excitation of He(2$^1$S$_0$) to He(2$^3$S$_1$) \cite{Phelps1955}. To test this assumption, we have also measured the pressure dependence of the helium singlet-to-triplet ratio (Fig. \ref{fig:stratio}). The results from this measurement clearly show that the ratio decreases as the valve stagnation pressure $p$ is increased, which suggests a higher rate of He(2$^1$S$_0$) $\rightarrow$ He(2$^3$S$_1$) conversion rate under these conditions.

Since the optical quenching efficiency does not depend on the absolute value of the singlet-to-triplet ratio, integrated signal intensities were used for the further analysis (integration range between 250--365 $\mu$s). Using Eq. \ref{eq:ratio}, an average singlet-to-triplet ratio $\bar{r}$ of 0.69 is obtained at $v = 1070$ m/s ($p = 10$ bar). The average singlet-to-triplet ratios at higher beam velocities (0.69 at $v = 1500$ m/s and $p = 15$ bar; 0.43 at $v = 1750$ m/s and $p = 30$ bar) were inferred by comparison with the results from numerical calculations (see discussion below). The different singlet-to-triplet ratios are attributed to changes in the valve characteristics and beam properties caused by the use of other valve stagnation pressures and valve temperatures.
\begin{figure*} 
	\includegraphics[width=\linewidth]{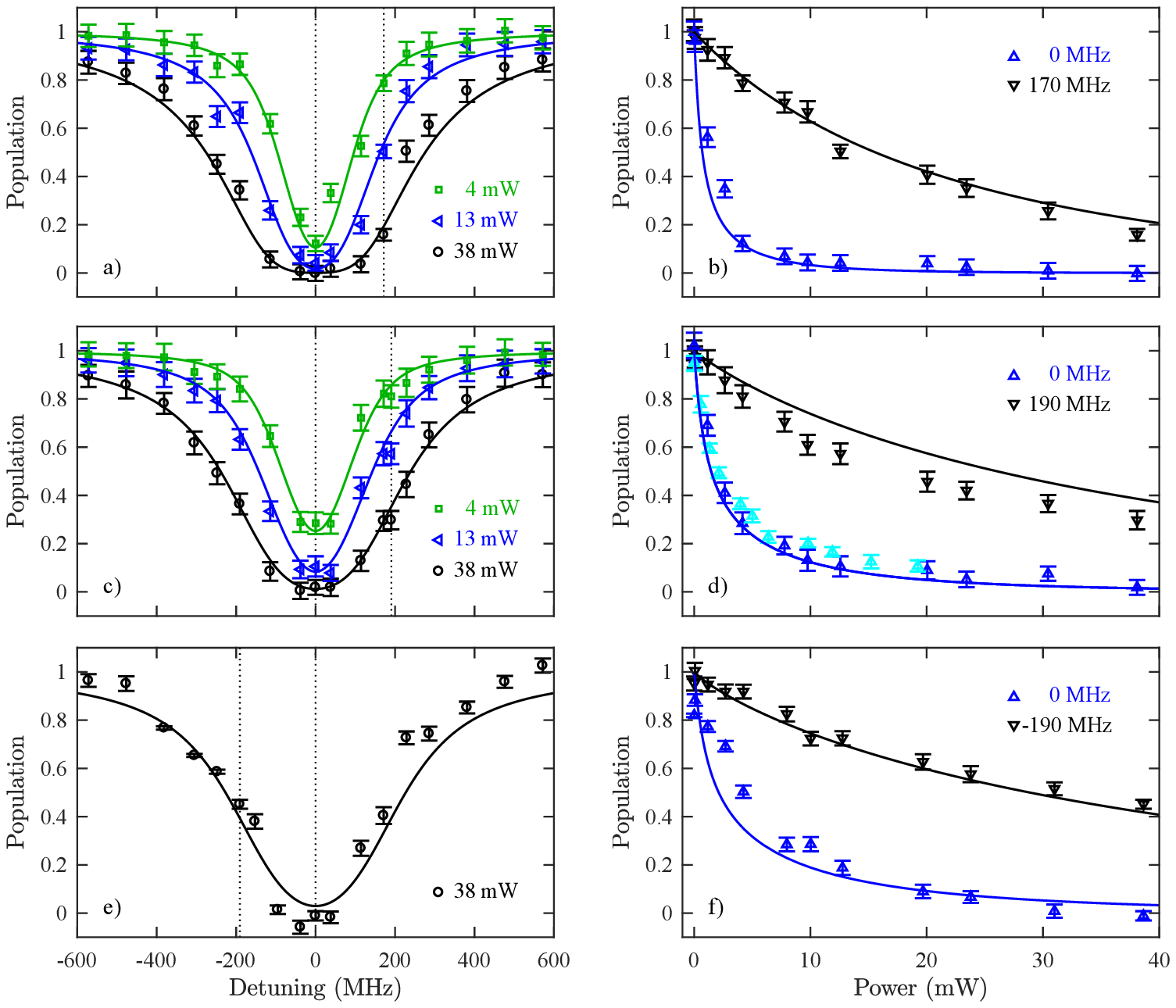}
	\caption{\label{fig:quenchresults} (a, c, e) He(2$^1$S$_0$) state population obtained from experimental measurements (markers) and from numerical calculations (solid lines) as a function of detuning from resonance at different laser powers (indicated in the legend). (b, d, f) Power-dependence of the He(2$^1$S$_0$) state populations obtained from experimental measurements (markers) and from numerical calculations (solid lines) at two different detunings from resonance (indicated in the legend). These laser detunings are also marked as vertical dashed lines in (a, c, e). The helium beam velocities were $v = 1070$ m/s (a, b), $v = 1500$ m/s (c, d), and $v = 1750$ m/s (e, f). The light-blue-colored markers in (d) are obtained from a measurement at zero detuning in which the back reflection of the laser light at mirror M6 was blocked. To allow for a comparison with the other measurements, the plotted laser power was divided by a factor of two.}
\end{figure*}

To quantify the optical depletion efficiency, measurements were done at various laser detunings, laser powers and at three different helium beam velocities. The experimental results were then converted to He(2$^1$S$_0$) state populations taking into account the average helium singlet-to-triplet ratio. From the experimental results at $v = 1070$ m/s at a laser power of 38 mW in Fig. \ref{fig:quenchresults} (a), it can be seen that the state population barely changes over a range of about 230 MHz around resonance. This is much larger than the natural linewidth of the transition ($\Delta \nu = 40$ MHz) and it is thus a clear indication that all of the He(2$^1$S$_0$) state population is depleted over a large range of detunings. We observe that the FWHM of the depletion curves in Fig. \ref{fig:quenchresults} (a-c) is reduced as the laser power is decreased, which is also consistent with the measured power dependence of the He(2$^1$S$_0$) state population in Fig. \ref{fig:quenchresults} (b). This figure illustrates that the depletion efficiency increases as a function of increasing laser power. On resonance, the population reaches a constant value at laser powers $\geq$ 10 mW which is another clear indication for complete population transfer out of the He(2$^1$S$_0$) state and it justifies the use of Eq. \ref{eq:ratio} for the determination of the singlet-to-triplet ratio at $v = 1070$ m/s. At a laser detuning of $\delta\nu =$ 170 MHz, the population does not reach a constant value, even at the highest laser power used in the experiment. At $v = 1500$ m/s and $v = 1750$ m/s, we observe similar wavelength and power dependences as for $v = 1070$ m/s. However, the quenching efficiency at low laser powers and at non-zero detunings is reduced owing to the increase in Doppler width at higher beam velocities (cf. Eq. \ref{eq:Doppler}).

To interpret the experimental results, a global fit procedure with two adjustable fit parameters was used for the numerical calculations. On the one hand, an effective orifice diameter $d_{\mathrm{eff}}$ was used to factor in the spatially broad distribution of metastable helium atoms which results from discharge excitation. As second parameter, an effective optical intensity $I_\mathrm{eff}$ was assumed to account for the inhomogeneous intensity distribution of the laser beam in the interaction zone. The inhomogeneous beam profile also results in a variation of the laser interaction time with the atoms in the supersonic beam. However, a change in laser interaction time is analogous to a variation of the laser intensity. Therefore, the laser interaction time was kept at a constant value of $t_{\mathrm{int}} = N\left(2a \right)/v$ at each helium beam velocity, where $N$ is the number of passes of the laser beam through the interaction volume. From a global fit to all the experimental datasets shown in Fig. \ref{fig:quenchresults}, we obtain $I_\mathrm{eff} = 0.004 \cdot A_\mathrm{l}/P_\mathrm{l}$ and $d_{\mathrm{eff}} = 48 d = 4.8$ mm. Given the approximations used for the model and for the fit, the results obtained from numerical calculations (shown as solid lines in Fig. \ref{fig:quenchresults}) are in good overall agreement with the experimental data.

The Doppler linewidths resulting from $d_{\mathrm{eff}} = 4.8$ mm are in between $\sigma_{\mathrm{D}} = 99$ MHz (at $v = 1070$ m/s) and 157 MHz (at $v = 1750$ m/s). The Doppler widths are thus much higher than the laser linewidth, which we estimate as $\sigma_{\mathrm{l}} \leq 20$ MHz from the FPI fringe pattern. The true laser linewidth is probably much smaller than that, but it could not be determined to a higher accuracy owing to the low finesse of the FPI. Therefore, the uncertainty of the laser linewidth does not significantly alter the outcome of the calculations. Deviations between theory and experiment can be attributed to the assumptions made in the numerical simulation, e.g. the use of a uniform laser intensity distribution. Experimental factors, such as small wavelength drifts inside the wavelength meter due to thermal fluctuations, can also not be ruled out. The use of a double-pass configuration (mirror M6 in Fig. \ref{fig:setup}) indeed leads to a two-fold increase of the interaction time, as can be seen from the experimental results of a single-pass measurement (light-blue markers in Fig. \ref{fig:quenchresults} (d)). Since the measurements were taken at different days, the deviation from the double-pass measurement (dark-blue markers in Fig. \ref{fig:quenchresults} (d)) can be related to changes in the helium singlet-to-triplet ratio or to a different alignment of the laser beam through the chamber.

The uncertainty of the singlet-to-triplet ratio provides a lower limit to the maximum quenching efficiency at each beam velocity. At $v = 1070$ m/s, the ratio can be directly inferred from Eq. \ref{eq:ratio} (see above) and the resulting quenching efficiency is thus known to a high accuracy. At $v = 1500$ m/s and at $v = 1750$ m/s, Eq. \ref{eq:ratio} may not hold owing to the decreased depletion efficiency (as a result of the increased Doppler width). In these cases, the average singlet-to-triplet ratios were determined by comparison with the results from numerical calculations. At zero detuning, unphysical negative He(2$^1$S$_0$) state populations would be obtained at the highest laser powers if the ratios were assumed to be 10 \% lower. If the ratios were 10 \% higher, the experimental results and the numerically calculated values would not agree. Therefore, the maximum on-resonance quenching efficiencies at a laser power of 38 mW are $100^{+1}_{-0}$ \% at $v = 1070$ m/s, $98^{+2}_{-5}$ \% at $v = 1500$ m/s and $97^{+3}_{-5}$ \% at $v = 1750$ m/s.

\section{Conclusion}
In this paper, we have described an original and very efficient technique for the optical depletion of He($2^1$S$_0$) state population in a supersonic beam. This scheme is comparably inexpensive and easily implemented using commercial or home-built diode laser systems. 
In the future, we are planning to use a laser lock based on saturated absorption spectroscopy. This will ensure a continuous on-resonance operation of the diode laser, and it will avoid the use of an expensive wavelength meter.

The optical depletion scheme will be especially beneficial for collision experiments. For example, in our laboratory, this setup will be used for the study of quantum-state-controlled reactive collisions between metastable helium atoms and lithium atoms. The optical quenching of He($2^1$S$_0$) will allow us to elucidate the relative contributions of the two metastable states of helium to the reaction, and it will thus make it possible to accurately describe the different reaction channels.

\section*{Acknowledgements}
We thank the group of S. Willitsch (University of Basel) and L. Petralia (University of Oxford) for technical advice on the diode laser design and on the implementation of the wavelength meter lock, respectively. This work is funded by the German Research Council (DFG) under projects DU1804/1-1 and GRK 2079. J. Grzesiak is thankful for additional financial support by the International Graduate Academy (IGA) of the Freiburg Research Services. K. Dulitz acknowledges support by the Chemical Industry Fund (FCI) through a Liebig Fellowship.


\begin{thebibliography}{46}%
	\makeatletter
	\providecommand \@ifxundefined [1]{%
		\@ifx{#1\undefined}
	}%
	\providecommand \@ifnum [1]{%
		\ifnum #1\expandafter \@firstoftwo
		\else \expandafter \@secondoftwo
		\fi
	}%
	\providecommand \@ifx [1]{%
		\ifx #1\expandafter \@firstoftwo
		\else \expandafter \@secondoftwo
		\fi
	}%
	\providecommand \natexlab [1]{#1}%
	\providecommand \enquote  [1]{``#1''}%
	\providecommand \bibnamefont  [1]{#1}%
	\providecommand \bibfnamefont [1]{#1}%
	\providecommand \citenamefont [1]{#1}%
	\providecommand \href@noop [0]{\@secondoftwo}%
	\providecommand \href [0]{\begingroup \@sanitize@url \@href}%
	\providecommand \@href[1]{\@@startlink{#1}\@@href}%
	\providecommand \@@href[1]{\endgroup#1\@@endlink}%
	\providecommand \@sanitize@url [0]{\catcode `\\12\catcode `\$12\catcode
		`\&12\catcode `\#12\catcode `\^12\catcode `\_12\catcode `\%12\relax}%
	\providecommand \@@startlink[1]{}%
	\providecommand \@@endlink[0]{}%
	\providecommand \url  [0]{\begingroup\@sanitize@url \@url }%
	\providecommand \@url [1]{\endgroup\@href {#1}{\urlprefix }}%
	\providecommand \urlprefix  [0]{URL }%
	\providecommand \Eprint [0]{\href }%
	\providecommand \doibase [0]{http://dx.doi.org/}%
	\providecommand \selectlanguage [0]{\@gobble}%
	\providecommand \bibinfo  [0]{\@secondoftwo}%
	\providecommand \bibfield  [0]{\@secondoftwo}%
	\providecommand \translation [1]{[#1]}%
	\providecommand \BibitemOpen [0]{}%
	\providecommand \bibitemStop [0]{}%
	\providecommand \bibitemNoStop [0]{.\EOS\space}%
	\providecommand \EOS [0]{\spacefactor3000\relax}%
	\providecommand \BibitemShut  [1]{\csname bibitem#1\endcsname}%
	\let\auto@bib@innerbib\@empty
	\bibitem [{\citenamefont {Makabe}\ \emph {et~al.}(1992)\citenamefont {Makabe},
		\citenamefont {Nakano},\ and\ \citenamefont {Yamaguchi}}]{Makabe1992}%
	\BibitemOpen
	\bibfield  {author} {\bibinfo {author} {\bibfnamefont {T.}~\bibnamefont
			{Makabe}}, \bibinfo {author} {\bibfnamefont {N.}~\bibnamefont {Nakano}}, \
		and\ \bibinfo {author} {\bibfnamefont {Y.}~\bibnamefont {Yamaguchi}},\
	}\bibfield  {title} {\enquote {\bibinfo {title} {{Modeling and diagnostics of
					the structure of rf glow discharges in Ar at 13.56 MHz}},}\ }\href {\doibase
		10.1103/PhysRevA.45.2520} {\bibfield  {journal} {\bibinfo  {journal} {Phys.
				Rev. A}\ }\textbf {\bibinfo {volume} {45}},\ \bibinfo {pages} {2520}
		(\bibinfo {year} {1992})}\BibitemShut {NoStop}%
	\bibitem [{\citenamefont {Falcinelli}\ \emph {et~al.}(2015)\citenamefont
		{Falcinelli}, \citenamefont {Pirani},\ and\ \citenamefont
		{Vecchiocattivi}}]{Falcinelli2015}%
	\BibitemOpen
	\bibfield  {author} {\bibinfo {author} {\bibfnamefont {S.}~\bibnamefont
			{Falcinelli}}, \bibinfo {author} {\bibfnamefont {F.}~\bibnamefont {Pirani}},
		\ and\ \bibinfo {author} {\bibfnamefont {F.}~\bibnamefont {Vecchiocattivi}},\
	}\bibfield  {title} {\enquote {\bibinfo {title} {{The possible role of
					Penning ionization processes in planetary atmospheres}},}\ }\href {\doibase
		10.3390/atmos6030299} {\bibfield  {journal} {\bibinfo  {journal}
			{Atmosphere}\ }\textbf {\bibinfo {volume} {6}},\ \bibinfo {pages} {299}
		(\bibinfo {year} {2015})}\BibitemShut {NoStop}%
	\bibitem [{\citenamefont {Gay}(1996)}]{Gay1996}%
	\BibitemOpen
	\bibfield  {author} {\bibinfo {author} {\bibfnamefont {T.~J.}\ \bibnamefont
			{Gay}},\ }\bibfield  {title} {\enquote {\bibinfo {title} {{6. Sources of
					metastable atoms and molecules}},}\ }in\ \href {\doibase
		10.1016/S0076-695X(08)60788-7} {\emph {\bibinfo {booktitle} {Atomic,
				molecular, and optical physics: atoms and molecules}}},\ \bibinfo {series}
	{Experimental methods in the physical sciences}, Vol.\ \bibinfo {volume} {29,
		Part B},\ \bibinfo {editor} {edited by\ \bibinfo {editor} {\bibfnamefont
			{F.}~\bibnamefont {Dunning}}\ and\ \bibinfo {editor} {\bibfnamefont {R.~G.}\
			\bibnamefont {Hulet}}}\ (\bibinfo  {publisher} {Academic Press},\ \bibinfo
	{year} {1996})\ pp.\ \bibinfo {pages} {95 -- 114}\BibitemShut {NoStop}%
	\bibitem [{\citenamefont {Harada}\ \emph {et~al.}(1997)\citenamefont {Harada},
		\citenamefont {Masuda},\ and\ \citenamefont {Ozaki}}]{Harada1997}%
	\BibitemOpen
	\bibfield  {author} {\bibinfo {author} {\bibfnamefont {Y.}~\bibnamefont
			{Harada}}, \bibinfo {author} {\bibfnamefont {S.}~\bibnamefont {Masuda}}, \
		and\ \bibinfo {author} {\bibfnamefont {H.}~\bibnamefont {Ozaki}},\ }\bibfield
	{title} {\enquote {\bibinfo {title} {{Electron spectroscopy using metastable
					atoms as probes for solid surfaces}},}\ }\href {\doibase 10.1021/cr940315v}
	{\bibfield  {journal} {\bibinfo  {journal} {Chem. Rev.}\ }\textbf {\bibinfo
			{volume} {97}},\ \bibinfo {pages} {1897} (\bibinfo {year}
		{1997})}\BibitemShut {NoStop}%
	\bibitem [{\citenamefont {Onellion}\ \emph {et~al.}(1984)\citenamefont
		{Onellion}, \citenamefont {Hart}, \citenamefont {Dunning},\ and\
		\citenamefont {Walters}}]{Onellion1984}%
	\BibitemOpen
	\bibfield  {author} {\bibinfo {author} {\bibfnamefont {M.}~\bibnamefont
			{Onellion}}, \bibinfo {author} {\bibfnamefont {M.~W.}\ \bibnamefont {Hart}},
		\bibinfo {author} {\bibfnamefont {F.~B.}\ \bibnamefont {Dunning}}, \ and\
		\bibinfo {author} {\bibfnamefont {G.~K.}\ \bibnamefont {Walters}},\
	}\bibfield  {title} {\enquote {\bibinfo {title} {{Spin-polarized
					metastable-atom deexcitation spectroscopy: a new probe of surface
					magnetism}},}\ }\href {\doibase 10.1103/PhysRevLett.52.380} {\bibfield
		{journal} {\bibinfo  {journal} {Phys. Rev. Lett.}\ }\textbf {\bibinfo
			{volume} {52}},\ \bibinfo {pages} {380} (\bibinfo {year} {1984})}\BibitemShut
	{NoStop}%
	\bibitem [{\citenamefont {Vassen}\ \emph {et~al.}(2012)\citenamefont {Vassen},
		\citenamefont {Cohen-Tannoudji}, \citenamefont {Leduc}, \citenamefont
		{Boiron}, \citenamefont {Westbrook}, \citenamefont {Truscott}, \citenamefont
		{Baldwin}, \citenamefont {Birkl}, \citenamefont {Cancio},\ and\ \citenamefont
		{Trippenbach}}]{Vassen2012}%
	\BibitemOpen
	\bibfield  {author} {\bibinfo {author} {\bibfnamefont {W.}~\bibnamefont
			{Vassen}}, \bibinfo {author} {\bibfnamefont {C.}~\bibnamefont
			{Cohen-Tannoudji}}, \bibinfo {author} {\bibfnamefont {M.}~\bibnamefont
			{Leduc}}, \bibinfo {author} {\bibfnamefont {D.}~\bibnamefont {Boiron}},
		\bibinfo {author} {\bibfnamefont {C.~I.}\ \bibnamefont {Westbrook}}, \bibinfo
		{author} {\bibfnamefont {A.}~\bibnamefont {Truscott}}, \bibinfo {author}
		{\bibfnamefont {K.}~\bibnamefont {Baldwin}}, \bibinfo {author} {\bibfnamefont
			{G.}~\bibnamefont {Birkl}}, \bibinfo {author} {\bibfnamefont
			{P.}~\bibnamefont {Cancio}}, \ and\ \bibinfo {author} {\bibfnamefont
			{M.}~\bibnamefont {Trippenbach}},\ }\bibfield  {title} {\enquote {\bibinfo
			{title} {{Cold and trapped metastable noble gases}},}\ }\href {\doibase
		10.1103/RevModPhys.84.175} {\bibfield  {journal} {\bibinfo  {journal} {Rev.
				Mod. Phys.}\ }\textbf {\bibinfo {volume} {84}},\ \bibinfo {pages} {175}
		(\bibinfo {year} {2012})}\BibitemShut {NoStop}%
	\bibitem [{\citenamefont {Vassen}\ \emph {et~al.}(2016)\citenamefont {Vassen},
		\citenamefont {Notermans}, \citenamefont {Rengelink},\ and\ \citenamefont
		{van~der Beek}}]{Vassen2016}%
	\BibitemOpen
	\bibfield  {author} {\bibinfo {author} {\bibfnamefont {W.}~\bibnamefont
			{Vassen}}, \bibinfo {author} {\bibfnamefont {R.~P. M. J.~W.}\ \bibnamefont
			{Notermans}}, \bibinfo {author} {\bibfnamefont {R.~J.}\ \bibnamefont
			{Rengelink}}, \ and\ \bibinfo {author} {\bibfnamefont {R.~F. H.~J.}\
			\bibnamefont {van~der Beek}},\ }\bibfield  {title} {\enquote {\bibinfo
			{title} {{Ultracold metastable helium: Ramsey fringes and atom
					interferometry}},}\ }\href {\doibase 10.1007/s00340-016-6563-0} {\bibfield
		{journal} {\bibinfo  {journal} {Appl. Phys. B}\ }\textbf {\bibinfo {volume}
			{122}},\ \bibinfo {pages} {289} (\bibinfo {year} {2016})}\BibitemShut
	{NoStop}%
	\bibitem [{\citenamefont {Aspect}\ \emph {et~al.}(1988)\citenamefont {Aspect},
		\citenamefont {Arimondo}, \citenamefont {Kaiser}, \citenamefont
		{Vansteenkiste},\ and\ \citenamefont {Cohen-Tannoudji}}]{Aspect1988}%
	\BibitemOpen
	\bibfield  {author} {\bibinfo {author} {\bibfnamefont {A.}~\bibnamefont
			{Aspect}}, \bibinfo {author} {\bibfnamefont {E.}~\bibnamefont {Arimondo}},
		\bibinfo {author} {\bibfnamefont {R.}~\bibnamefont {Kaiser}}, \bibinfo
		{author} {\bibfnamefont {N.}~\bibnamefont {Vansteenkiste}}, \ and\ \bibinfo
		{author} {\bibfnamefont {C.}~\bibnamefont {Cohen-Tannoudji}},\ }\bibfield
	{title} {\enquote {\bibinfo {title} {{Laser cooling below the one-photon
					recoil energy by velocity-selective coherent population trapping}},}\ }\href
	{\doibase 10.1103/PhysRevLett.61.826} {\bibfield  {journal} {\bibinfo
			{journal} {Phys. Rev. Lett.}\ }\textbf {\bibinfo {volume} {61}},\ \bibinfo
		{pages} {826} (\bibinfo {year} {1988})}\BibitemShut {NoStop}%
	\bibitem [{\citenamefont {Hack}\ \emph {et~al.}(2000)\citenamefont {Hack},
		\citenamefont {Liu}, \citenamefont {Olshanii},\ and\ \citenamefont
		{Metcalf}}]{Hack2000}%
	\BibitemOpen
	\bibfield  {author} {\bibinfo {author} {\bibfnamefont {J.}~\bibnamefont
			{Hack}}, \bibinfo {author} {\bibfnamefont {L.}~\bibnamefont {Liu}}, \bibinfo
		{author} {\bibfnamefont {M.}~\bibnamefont {Olshanii}}, \ and\ \bibinfo
		{author} {\bibfnamefont {H.}~\bibnamefont {Metcalf}},\ }\bibfield  {title}
	{\enquote {\bibinfo {title} {{Velocity-selective coherent population trapping
					of two-level atoms}},}\ }\href {\doibase 10.1103/PhysRevA.62.013405}
	{\bibfield  {journal} {\bibinfo  {journal} {Phys. Rev. A}\ }\textbf {\bibinfo
			{volume} {62}},\ \bibinfo {pages} {013405} (\bibinfo {year}
		{2000})}\BibitemShut {NoStop}%
	\bibitem [{\citenamefont {Rasel}\ \emph {et~al.}(1999)\citenamefont {Rasel},
		\citenamefont {Pereira Dos~Santos}, \citenamefont {Saverio~Pavone},
		\citenamefont {Perales}, \citenamefont {Unnikrishnan},\ and\ \citenamefont
		{Leduc}}]{Rasel1999}%
	\BibitemOpen
	\bibfield  {author} {\bibinfo {author} {\bibfnamefont {E.}~\bibnamefont
			{Rasel}}, \bibinfo {author} {\bibfnamefont {F.}~\bibnamefont {Pereira
				Dos~Santos}}, \bibinfo {author} {\bibfnamefont {F.}~\bibnamefont
			{Saverio~Pavone}}, \bibinfo {author} {\bibfnamefont {F.}~\bibnamefont
			{Perales}}, \bibinfo {author} {\bibfnamefont {C.}~\bibnamefont
			{Unnikrishnan}}, \ and\ \bibinfo {author} {\bibfnamefont {M.}~\bibnamefont
			{Leduc}},\ }\bibfield  {title} {\enquote {\bibinfo {title} {{White light
					transverse cooling of a helium beam}},}\ }\href {\doibase
		10.1007/s100530050573} {\bibfield  {journal} {\bibinfo  {journal} {Eur. Phys.
				J. D}\ }\textbf {\bibinfo {volume} {7}},\ \bibinfo {pages} {311} (\bibinfo
		{year} {1999})}\BibitemShut {NoStop}%
	\bibitem [{\citenamefont {Cashen}\ and\ \citenamefont
		{Metcalf}(2001)}]{Cashen2001}%
	\BibitemOpen
	\bibfield  {author} {\bibinfo {author} {\bibfnamefont {M.~T.}\ \bibnamefont
			{Cashen}}\ and\ \bibinfo {author} {\bibfnamefont {H.}~\bibnamefont
			{Metcalf}},\ }\bibfield  {title} {\enquote {\bibinfo {title} {{Bichromatic
					force on helium}},}\ }\href {\doibase 10.1103/PhysRevA.63.025406} {\bibfield
		{journal} {\bibinfo  {journal} {Phys. Rev. A}\ }\textbf {\bibinfo {volume}
			{63}},\ \bibinfo {pages} {025406} (\bibinfo {year} {2001})}\BibitemShut
	{NoStop}%
	\bibitem [{\citenamefont {Partlow}\ \emph {et~al.}(2004)\citenamefont
		{Partlow}, \citenamefont {Miao}, \citenamefont {Bochmann}, \citenamefont
		{Cashen},\ and\ \citenamefont {Metcalf}}]{Partlow2004}%
	\BibitemOpen
	\bibfield  {author} {\bibinfo {author} {\bibfnamefont {M.}~\bibnamefont
			{Partlow}}, \bibinfo {author} {\bibfnamefont {X.}~\bibnamefont {Miao}},
		\bibinfo {author} {\bibfnamefont {J.}~\bibnamefont {Bochmann}}, \bibinfo
		{author} {\bibfnamefont {M.}~\bibnamefont {Cashen}}, \ and\ \bibinfo {author}
		{\bibfnamefont {H.}~\bibnamefont {Metcalf}},\ }\bibfield  {title} {\enquote
		{\bibinfo {title} {{Bichromatic slowing and collimation to make an intense
					helium beam}},}\ }\href {\doibase 10.1103/PhysRevLett.93.213004} {\bibfield
		{journal} {\bibinfo  {journal} {Phys. Rev. Lett.}\ }\textbf {\bibinfo
			{volume} {93}},\ \bibinfo {pages} {213004} (\bibinfo {year}
		{2004})}\BibitemShut {NoStop}%
	\bibitem [{\citenamefont {Corder}\ \emph {et~al.}(2015)\citenamefont {Corder},
		\citenamefont {Arnold},\ and\ \citenamefont {Metcalf}}]{Corder2015}%
	\BibitemOpen
	\bibfield  {author} {\bibinfo {author} {\bibfnamefont {C.}~\bibnamefont
			{Corder}}, \bibinfo {author} {\bibfnamefont {B.}~\bibnamefont {Arnold}}, \
		and\ \bibinfo {author} {\bibfnamefont {H.}~\bibnamefont {Metcalf}},\
	}\bibfield  {title} {\enquote {\bibinfo {title} {{Laser cooling without
					spontaneous emission}},}\ }\href {\doibase 10.1103/PhysRevLett.114.043002}
	{\bibfield  {journal} {\bibinfo  {journal} {Phys. Rev. Lett.}\ }\textbf
		{\bibinfo {volume} {114}},\ \bibinfo {pages} {043002} (\bibinfo {year}
		{2015})}\BibitemShut {NoStop}%
	\bibitem [{\citenamefont {Dulitz}\ \emph {et~al.}(2015)\citenamefont {Dulitz},
		\citenamefont {Tauschinsky},\ and\ \citenamefont {Softley}}]{Dulitz2015a}%
	\BibitemOpen
	\bibfield  {author} {\bibinfo {author} {\bibfnamefont {K.}~\bibnamefont
			{Dulitz}}, \bibinfo {author} {\bibfnamefont {A.}~\bibnamefont {Tauschinsky}},
		\ and\ \bibinfo {author} {\bibfnamefont {T.~P.}\ \bibnamefont {Softley}},\
	}\bibfield  {title} {\enquote {\bibinfo {title} {{Zeeman deceleration of
					electron-impact-excited metastable helium atoms}},}\ }\href
	{http://stacks.iop.org/1367-2630/17/i=3/a=035005} {\bibfield  {journal}
		{\bibinfo  {journal} {New J. Phys.}\ }\textbf {\bibinfo {volume} {17}},\
		\bibinfo {pages} {035005} (\bibinfo {year} {2015})}\BibitemShut {NoStop}%
	\bibitem [{\citenamefont {Siska}(1993)}]{Siska1993}%
	\BibitemOpen
	\bibfield  {author} {\bibinfo {author} {\bibfnamefont {P.~E.}\ \bibnamefont
			{Siska}},\ }\bibfield  {title} {\enquote {\bibinfo {title} {{Molecular-beam
					studies of Penning ionization}},}\ }\href {\doibase
		10.1103/RevModPhys.65.337} {\bibfield  {journal} {\bibinfo  {journal} {Rev.
				Mod. Phys.}\ }\textbf {\bibinfo {volume} {65}},\ \bibinfo {pages} {337}
		(\bibinfo {year} {1993})}\BibitemShut {NoStop}%
	\bibitem [{\citenamefont {Henson}\ \emph {et~al.}(2012)\citenamefont {Henson},
		\citenamefont {Gersten}, \citenamefont {Shagam}, \citenamefont {Narevicius},\
		and\ \citenamefont {Narevicius}}]{Henson2012}%
	\BibitemOpen
	\bibfield  {author} {\bibinfo {author} {\bibfnamefont {A.~B.}\ \bibnamefont
			{Henson}}, \bibinfo {author} {\bibfnamefont {S.}~\bibnamefont {Gersten}},
		\bibinfo {author} {\bibfnamefont {Y.}~\bibnamefont {Shagam}}, \bibinfo
		{author} {\bibfnamefont {J.}~\bibnamefont {Narevicius}}, \ and\ \bibinfo
		{author} {\bibfnamefont {E.}~\bibnamefont {Narevicius}},\ }\bibfield  {title}
	{\enquote {\bibinfo {title} {{Observation of resonances in Penning ionization
					reactions at sub-Kelvin temperatures in merged beams}},}\ }\href {\doibase
		10.1126/science.1229141} {\bibfield  {journal} {\bibinfo  {journal}
			{Science}\ }\textbf {\bibinfo {volume} {338}},\ \bibinfo {pages} {234}
		(\bibinfo {year} {2012})}\BibitemShut {NoStop}%
	\bibitem [{\citenamefont {Gordon}\ \emph {et~al.}(2017)\citenamefont {Gordon},
		\citenamefont {Zou}, \citenamefont {Tanteri}, \citenamefont {Jankunas},\ and\
		\citenamefont {Osterwalder}}]{Gordon2017}%
	\BibitemOpen
	\bibfield  {author} {\bibinfo {author} {\bibfnamefont {S.~D.~S.}\
			\bibnamefont {Gordon}}, \bibinfo {author} {\bibfnamefont {J.}~\bibnamefont
			{Zou}}, \bibinfo {author} {\bibfnamefont {S.}~\bibnamefont {Tanteri}},
		\bibinfo {author} {\bibfnamefont {J.}~\bibnamefont {Jankunas}}, \ and\
		\bibinfo {author} {\bibfnamefont {A.}~\bibnamefont {Osterwalder}},\
	}\bibfield  {title} {\enquote {\bibinfo {title} {{Energy dependent
					stereodynamics of the $\mathrm{Ne}({^{3}\mathrm{P}}_{2})+\mathrm{Ar}$
					reaction}},}\ }\href {\doibase 10.1103/PhysRevLett.119.053001} {\bibfield
		{journal} {\bibinfo  {journal} {Phys. Rev. Lett.}\ }\textbf {\bibinfo
			{volume} {119}},\ \bibinfo {pages} {053001} (\bibinfo {year}
		{2017})}\BibitemShut {NoStop}%
	\bibitem [{\citenamefont {Fedichev}\ \emph {et~al.}(1996)\citenamefont
		{Fedichev}, \citenamefont {Reynolds}, \citenamefont {Rahmanov},\ and\
		\citenamefont {Shlyapnikov}}]{Fedichev1996}%
	\BibitemOpen
	\bibfield  {author} {\bibinfo {author} {\bibfnamefont {P.~O.}\ \bibnamefont
			{Fedichev}}, \bibinfo {author} {\bibfnamefont {M.~W.}\ \bibnamefont
			{Reynolds}}, \bibinfo {author} {\bibfnamefont {U.~M.}\ \bibnamefont
			{Rahmanov}}, \ and\ \bibinfo {author} {\bibfnamefont {G.~V.}\ \bibnamefont
			{Shlyapnikov}},\ }\bibfield  {title} {\enquote {\bibinfo {title} {{Inelastic
					decay processes in a gas of spin-polarized triplet helium}},}\ }\href
	{\doibase 10.1103/PhysRevA.53.1447} {\bibfield  {journal} {\bibinfo
			{journal} {Phys. Rev. A}\ }\textbf {\bibinfo {volume} {53}},\ \bibinfo
		{pages} {1447} (\bibinfo {year} {1996})}\BibitemShut {NoStop}%
	\bibitem [{\citenamefont {Herschbach}\ \emph {et~al.}(2000)\citenamefont
		{Herschbach}, \citenamefont {Tol}, \citenamefont {Hogervorst},\ and\
		\citenamefont {Vassen}}]{Herschbach2000}%
	\BibitemOpen
	\bibfield  {author} {\bibinfo {author} {\bibfnamefont {N.}~\bibnamefont
			{Herschbach}}, \bibinfo {author} {\bibfnamefont {P.~J.~J.}\ \bibnamefont
			{Tol}}, \bibinfo {author} {\bibfnamefont {W.}~\bibnamefont {Hogervorst}}, \
		and\ \bibinfo {author} {\bibfnamefont {W.}~\bibnamefont {Vassen}},\
	}\bibfield  {title} {\enquote {\bibinfo {title} {{Suppression of Penning
					ionization by spin polarization of cold $\mathrm{He}(2{}^{3}\mathrm{S})$
					atoms}},}\ }\href {\doibase 10.1103/PhysRevA.61.050702} {\bibfield  {journal}
		{\bibinfo  {journal} {Phys. Rev. A}\ }\textbf {\bibinfo {volume} {61}},\
		\bibinfo {pages} {050702} (\bibinfo {year} {2000})}\BibitemShut {NoStop}%
	\bibitem [{\citenamefont {Rundel}\ \emph {et~al.}(1974)\citenamefont {Rundel},
		\citenamefont {Dunning},\ and\ \citenamefont {Stebbings}}]{Rundel1974a}%
	\BibitemOpen
	\bibfield  {author} {\bibinfo {author} {\bibfnamefont {R.~D.}\ \bibnamefont
			{Rundel}}, \bibinfo {author} {\bibfnamefont {F.~B.}\ \bibnamefont {Dunning}},
		\ and\ \bibinfo {author} {\bibfnamefont {R.~F.}\ \bibnamefont {Stebbings}},\
	}\bibfield  {title} {\enquote {\bibinfo {title} {{Velocity distributions in
					metastable atom beams produced by coaxial electron impact}},}\ }\href
	{\doibase http://dx.doi.org/10.1063/1.1686422} {\bibfield  {journal}
		{\bibinfo  {journal} {Rev. Sci. Instrum.}\ }\textbf {\bibinfo {volume}
			{45}},\ \bibinfo {pages} {116} (\bibinfo {year} {1974})}\BibitemShut
	{NoStop}%
	\bibitem [{\citenamefont {Dunning}\ \emph {et~al.}(1975)\citenamefont
		{Dunning}, \citenamefont {Cook}, \citenamefont {West},\ and\ \citenamefont
		{Stebbings}}]{Dunning1975}%
	\BibitemOpen
	\bibfield  {author} {\bibinfo {author} {\bibfnamefont {F.~B.}\ \bibnamefont
			{Dunning}}, \bibinfo {author} {\bibfnamefont {T.~B.}\ \bibnamefont {Cook}},
		\bibinfo {author} {\bibfnamefont {W.~P.}\ \bibnamefont {West}}, \ and\
		\bibinfo {author} {\bibfnamefont {R.~F.}\ \bibnamefont {Stebbings}},\
	}\bibfield  {title} {\enquote {\bibinfo {title} {{Selective removal of either
					metastable species from a mixed $^{3}\mathrm{P}_{0,2}$ rare-gas metastable
					beam}},}\ }\href {\doibase http://dx.doi.org/10.1063/1.1134403} {\bibfield
		{journal} {\bibinfo  {journal} {Rev. Sci. Instrum.}\ }\textbf {\bibinfo
			{volume} {46}},\ \bibinfo {pages} {1072} (\bibinfo {year}
		{1975})}\BibitemShut {NoStop}%
	\bibitem [{\citenamefont {Weiser}\ and\ \citenamefont
		{Siska}(1987)}]{Weiser1987}%
	\BibitemOpen
	\bibfield  {author} {\bibinfo {author} {\bibfnamefont {C.}~\bibnamefont
			{Weiser}}\ and\ \bibinfo {author} {\bibfnamefont {P.~E.}\ \bibnamefont
			{Siska}},\ }\bibfield  {title} {\enquote {\bibinfo {title} {{Magnetic
					deflection analysis of supersonic metastable atom beams}},}\ }\href {\doibase
		10.1063/1.1139474} {\bibfield  {journal} {\bibinfo  {journal} {Rev. Sci.
				Instrum.}\ }\textbf {\bibinfo {volume} {58}},\ \bibinfo {pages} {2124}
		(\bibinfo {year} {1987})}\BibitemShut {NoStop}%
	\bibitem [{\citenamefont {Theuer}\ and\ \citenamefont
		{Bergmann}(1998)}]{Theuer1998}%
	\BibitemOpen
	\bibfield  {author} {\bibinfo {author} {\bibfnamefont {H.}~\bibnamefont
			{Theuer}}\ and\ \bibinfo {author} {\bibfnamefont {K.}~\bibnamefont
			{Bergmann}},\ }\bibfield  {title} {\enquote {\bibinfo {title} {{Atomic beam
					deflection by coherent momentum transfer and the dependence on weak magnetic
					fields}},}\ }\href {\doibase 10.1007/s100530050141} {\bibfield  {journal}
		{\bibinfo  {journal} {Eur. Phys. J. D}\ }\textbf {\bibinfo {volume} {2}},\
		\bibinfo {pages} {279} (\bibinfo {year} {1998})}\BibitemShut {NoStop}%
	\bibitem [{\citenamefont {Aspect}\ \emph {et~al.}(1990)\citenamefont {Aspect},
		\citenamefont {Vansteenkiste}, \citenamefont {Kaiser}, \citenamefont
		{Haberland},\ and\ \citenamefont {Karrais}}]{Aspect1990}%
	\BibitemOpen
	\bibfield  {author} {\bibinfo {author} {\bibfnamefont {A.}~\bibnamefont
			{Aspect}}, \bibinfo {author} {\bibfnamefont {N.}~\bibnamefont
			{Vansteenkiste}}, \bibinfo {author} {\bibfnamefont {R.}~\bibnamefont
			{Kaiser}}, \bibinfo {author} {\bibfnamefont {H.}~\bibnamefont {Haberland}}, \
		and\ \bibinfo {author} {\bibfnamefont {M.}~\bibnamefont {Karrais}},\
	}\bibfield  {title} {\enquote {\bibinfo {title} {{Preparation of a pure
					intense beam of metastable helium by laser cooling}},}\ }\href {\doibase
		http://dx.doi.org/10.1016/0301-0104(90)89122-7} {\bibfield  {journal}
		{\bibinfo  {journal} {Chem. Phys.}\ }\textbf {\bibinfo {volume} {145}},\
		\bibinfo {pages} {307 } (\bibinfo {year} {1990})}\BibitemShut {NoStop}%
	\bibitem [{\citenamefont {Fry}\ and\ \citenamefont {Williams}(1969)}]{Fry1969}%
	\BibitemOpen
	\bibfield  {author} {\bibinfo {author} {\bibfnamefont {E.~S.}\ \bibnamefont
			{Fry}}\ and\ \bibinfo {author} {\bibfnamefont {W.~L.}\ \bibnamefont
			{Williams}},\ }\bibfield  {title} {\enquote {\bibinfo {title} {{Production of
					a $2^{3}\mathrm{S}$ helium beam}},}\ }\href {\doibase
		http://dx.doi.org/10.1063/1.1684181} {\bibfield  {journal} {\bibinfo
			{journal} {Rev. Sci. Instrum.}\ }\textbf {\bibinfo {volume} {40}},\ \bibinfo
		{pages} {1141} (\bibinfo {year} {1969})}\BibitemShut {NoStop}%
	\bibitem [{\citenamefont {Hotop}\ \emph {et~al.}(1969)\citenamefont {Hotop},
		\citenamefont {Niehaus},\ and\ \citenamefont {Schmeltekopf}}]{Hotop1969a}%
	\BibitemOpen
	\bibfield  {author} {\bibinfo {author} {\bibfnamefont {H.}~\bibnamefont
			{Hotop}}, \bibinfo {author} {\bibfnamefont {A.}~\bibnamefont {Niehaus}}, \
		and\ \bibinfo {author} {\bibfnamefont {A.~L.}\ \bibnamefont {Schmeltekopf}},\
	}\bibfield  {title} {\enquote {\bibinfo {title} {{Reactions of excited atoms
					and molecules with atoms and molecules}},}\ }\href {\doibase
		10.1007/BF01394439} {\bibfield  {journal} {\bibinfo  {journal} {Z. Phys. A -
				Hadron Nucl.}\ }\textbf {\bibinfo {volume} {229}},\ \bibinfo {pages} {1}
		(\bibinfo {year} {1969})}\BibitemShut {NoStop}%
	\bibitem [{\citenamefont {Hotop}\ \emph {et~al.}(1981)\citenamefont {Hotop},
		\citenamefont {Lorenzen},\ and\ \citenamefont {Zastrow}}]{Hotop1981}%
	\BibitemOpen
	\bibfield  {author} {\bibinfo {author} {\bibfnamefont {H.}~\bibnamefont
			{Hotop}}, \bibinfo {author} {\bibfnamefont {J.}~\bibnamefont {Lorenzen}}, \
		and\ \bibinfo {author} {\bibfnamefont {A.}~\bibnamefont {Zastrow}},\
	}\bibfield  {title} {\enquote {\bibinfo {title} {{Penning ionization electron
					spectrometry with state-selected, thermal-energy neon metastable atoms:
					Ne($^{3}\mathrm{P}_{2}$), Ne($^{3}\mathrm{P}_{0}$) + Ar, Kr, Xe, Hg}},}\
	}\href {\doibase http://dx.doi.org/10.1016/0368-2048(81)85045-1} {\bibfield
		{journal} {\bibinfo  {journal} {J. Electron Spectrosc.}\ }\textbf {\bibinfo
			{volume} {23}},\ \bibinfo {pages} {347 } (\bibinfo {year}
		{1981})}\BibitemShut {NoStop}%
	\bibitem [{\citenamefont {Harada}\ and\ \citenamefont
		{Ozaki}(1987)}]{Harada1987}%
	\BibitemOpen
	\bibfield  {author} {\bibinfo {author} {\bibfnamefont {Y.}~\bibnamefont
			{Harada}}\ and\ \bibinfo {author} {\bibfnamefont {H.}~\bibnamefont {Ozaki}},\
	}\bibfield  {title} {\enquote {\bibinfo {title} {{Penning ionization electron
					spectoscopy: its application to surface characterization of organic
					solids}},}\ }\href {http://stacks.iop.org/1347-4065/26/i=8R/a=1201}
	{\bibfield  {journal} {\bibinfo  {journal} {Jpn. J. Appl. Phys.}\ }\textbf
		{\bibinfo {volume} {26}},\ \bibinfo {pages} {1201} (\bibinfo {year}
		{1987})}\BibitemShut {NoStop}%
	\bibitem [{\citenamefont {Gaily}\ \emph {et~al.}(1980)\citenamefont {Gaily},
		\citenamefont {Coggiola}, \citenamefont {Peterson},\ and\ \citenamefont
		{Gillen}}]{Gaily1980}%
	\BibitemOpen
	\bibfield  {author} {\bibinfo {author} {\bibfnamefont {T.~D.}\ \bibnamefont
			{Gaily}}, \bibinfo {author} {\bibfnamefont {M.~J.}\ \bibnamefont {Coggiola}},
		\bibinfo {author} {\bibfnamefont {J.~R.}\ \bibnamefont {Peterson}}, \ and\
		\bibinfo {author} {\bibfnamefont {K.~T.}\ \bibnamefont {Gillen}},\ }\bibfield
	{title} {\enquote {\bibinfo {title} {{State purification of a fast neon
					metastable beam by collinear optical pumping}},}\ }\href {\doibase
		10.1063/1.1136400} {\bibfield  {journal} {\bibinfo  {journal} {Rev. Sci.
				Instrum.}\ }\textbf {\bibinfo {volume} {51}},\ \bibinfo {pages} {1168}
		(\bibinfo {year} {1980})}\BibitemShut {NoStop}%
	\bibitem [{\citenamefont {Verheijen}\ and\ \citenamefont
		{Beijerinck}(1986)}]{Verheijen1986}%
	\BibitemOpen
	\bibfield  {author} {\bibinfo {author} {\bibfnamefont {M.~J.}\ \bibnamefont
			{Verheijen}}\ and\ \bibinfo {author} {\bibfnamefont {H.~C.~W.}\ \bibnamefont
			{Beijerinck}},\ }\bibfield  {title} {\enquote {\bibinfo {title} {{State
					selected total Penning ionisation cross sections for the systems
					Ne*($^{3}\mathrm{P}_{0}$, $^{3}\mathrm{P}_{2}$) + Ar, Kr, Xe and N$_2$ in the
					energy range 0.06$<E_0\mathrm{(eV)}<$8.0}},}\ }\href {\doibase
		10.1016/0301-0104(86)85136-9} {\bibfield  {journal} {\bibinfo  {journal}
			{Chem. Phys.}\ }\textbf {\bibinfo {volume} {102}},\ \bibinfo {pages} {255}
		(\bibinfo {year} {1986})}\BibitemShut {NoStop}%
	\bibitem [{\citenamefont {Brand}\ \emph {et~al.}(1992)\citenamefont {Brand},
		\citenamefont {Furst}, \citenamefont {Gay},\ and\ \citenamefont
		{Schearer}}]{Brand1992}%
	\BibitemOpen
	\bibfield  {author} {\bibinfo {author} {\bibfnamefont {J.~A.}\ \bibnamefont
			{Brand}}, \bibinfo {author} {\bibfnamefont {J.~E.}\ \bibnamefont {Furst}},
		\bibinfo {author} {\bibfnamefont {T.~J.}\ \bibnamefont {Gay}}, \ and\
		\bibinfo {author} {\bibfnamefont {L.~D.}\ \bibnamefont {Schearer}},\
	}\bibfield  {title} {\enquote {\bibinfo {title} {{Production of a
					high-density state-selected metastable neon beam}},}\ }\href {\doibase
		10.1063/1.1143000} {\bibfield  {journal} {\bibinfo  {journal} {Rev. Sci.
				Instrum.}\ }\textbf {\bibinfo {volume} {63}},\ \bibinfo {pages} {163}
		(\bibinfo {year} {1992})}\BibitemShut {NoStop}%
	\bibitem [{\citenamefont {Kau}\ \emph {et~al.}(1998)\citenamefont {Kau},
		\citenamefont {Petrov}, \citenamefont {Sukhorukov},\ and\ \citenamefont
		{Hotop}}]{Kau1998}%
	\BibitemOpen
	\bibfield  {author} {\bibinfo {author} {\bibfnamefont {R.}~\bibnamefont
			{Kau}}, \bibinfo {author} {\bibfnamefont {I.~D.}\ \bibnamefont {Petrov}},
		\bibinfo {author} {\bibfnamefont {V.~L.}\ \bibnamefont {Sukhorukov}}, \ and\
		\bibinfo {author} {\bibfnamefont {H.}~\bibnamefont {Hotop}},\ }\bibfield
	{title} {\enquote {\bibinfo {title} {{Experimental and theoretical cross
					sections for photoionization of metastable Ar* and Kr* atoms near
					threshold}},}\ }\href {http://stacks.iop.org/0953-4075/31/i=5/a=010}
	{\bibfield  {journal} {\bibinfo  {journal} {J. Phys. B: At. Mol. Phys.}\
		}\textbf {\bibinfo {volume} {31}},\ \bibinfo {pages} {1011} (\bibinfo {year}
		{1998})}\BibitemShut {NoStop}%
	\bibitem [{\citenamefont {Thiel}\ \emph {et~al.}(2004)\citenamefont {Thiel},
		\citenamefont {Thiel}, \citenamefont {Yencha}, \citenamefont {Ruf},
		\citenamefont {Meyer},\ and\ \citenamefont {Hotop}}]{Thiel2004}%
	\BibitemOpen
	\bibfield  {author} {\bibinfo {author} {\bibfnamefont {F.~A.~U.}\
			\bibnamefont {Thiel}}, \bibinfo {author} {\bibfnamefont {L.}~\bibnamefont
			{Thiel}}, \bibinfo {author} {\bibfnamefont {A.~J.}\ \bibnamefont {Yencha}},
		\bibinfo {author} {\bibfnamefont {M.-W.}\ \bibnamefont {Ruf}}, \bibinfo
		{author} {\bibfnamefont {W.}~\bibnamefont {Meyer}}, \ and\ \bibinfo {author}
		{\bibfnamefont {H.}~\bibnamefont {Hotop}},\ }\bibfield  {title} {\enquote
		{\bibinfo {title} {{Experimental and theoretical electron energy spectra due
					to ionizing collisions of metastable Ar*($^{3}\mathrm{P}_{2}$),
					Ar*($^{3}\mathrm{P}_{0}$) and Kr*($^{3}\mathrm{P}_{0}$) atoms with
					ground-state Hg atoms}},}\ }\href
	{http://stacks.iop.org/0953-4075/37/i=18/a=010} {\bibfield  {journal}
		{\bibinfo  {journal} {J. Phys. B: At. Mol. Phys.}\ }\textbf {\bibinfo
			{volume} {37}},\ \bibinfo {pages} {3691} (\bibinfo {year}
		{2004})}\BibitemShut {NoStop}%
	\bibitem [{\citenamefont {Kato}\ \emph {et~al.}(2012)\citenamefont {Kato},
		\citenamefont {Fitzakerley}, \citenamefont {George}, \citenamefont {Vutha},
		\citenamefont {Weel}, \citenamefont {Storry}, \citenamefont {Kirchner},\ and\
		\citenamefont {Hessels}}]{Kato2012}%
	\BibitemOpen
	\bibfield  {author} {\bibinfo {author} {\bibfnamefont {K.}~\bibnamefont
			{Kato}}, \bibinfo {author} {\bibfnamefont {D.~W.}\ \bibnamefont
			{Fitzakerley}}, \bibinfo {author} {\bibfnamefont {M.~C.}\ \bibnamefont
			{George}}, \bibinfo {author} {\bibfnamefont {A.~C.}\ \bibnamefont {Vutha}},
		\bibinfo {author} {\bibfnamefont {M.}~\bibnamefont {Weel}}, \bibinfo {author}
		{\bibfnamefont {C.~H.}\ \bibnamefont {Storry}}, \bibinfo {author}
		{\bibfnamefont {T.}~\bibnamefont {Kirchner}}, \ and\ \bibinfo {author}
		{\bibfnamefont {E.~A.}\ \bibnamefont {Hessels}},\ }\bibfield  {title}
	{\enquote {\bibinfo {title} {{Selective detection of metastable helium atoms
					by elastic scattering collisions}},}\ }\href {\doibase
		10.1103/PhysRevA.86.014702} {\bibfield  {journal} {\bibinfo  {journal} {Phys.
				Rev. A}\ }\textbf {\bibinfo {volume} {86}},\ \bibinfo {pages} {014702}
		(\bibinfo {year} {2012})}\BibitemShut {NoStop}%
	\bibitem [{\citenamefont {Martin}(1960)}]{Martin1960}%
	\BibitemOpen
	\bibfield  {author} {\bibinfo {author} {\bibfnamefont {W.~C.}\ \bibnamefont
			{Martin}},\ }\bibfield  {title} {\enquote {\bibinfo {title} {{New wavelengths
					for some helium (He I) lines}},}\ }\href {\doibase 10.1364/JOSA.50.000174}
	{\bibfield  {journal} {\bibinfo  {journal} {J. Opt. Soc. Am.}\ }\textbf
		{\bibinfo {volume} {50}},\ \bibinfo {pages} {174} (\bibinfo {year}
		{1960})}\BibitemShut {NoStop}%
	\bibitem [{\citenamefont {Grzesiak}\ \emph {et~al.}(2018)\citenamefont
		{Grzesiak}, \citenamefont {Vashishta}, \citenamefont {Djuricanin},
		\citenamefont {Stienkemeier}, \citenamefont {Mudrich}, \citenamefont
		{Dulitz},\ and\ \citenamefont {Momose}}]{Grzesiak2018}%
	\BibitemOpen
	\bibfield  {author} {\bibinfo {author} {\bibfnamefont {J.}~\bibnamefont
			{Grzesiak}}, \bibinfo {author} {\bibfnamefont {M.}~\bibnamefont {Vashishta}},
		\bibinfo {author} {\bibfnamefont {P.}~\bibnamefont {Djuricanin}}, \bibinfo
		{author} {\bibfnamefont {F.}~\bibnamefont {Stienkemeier}}, \bibinfo {author}
		{\bibfnamefont {M.}~\bibnamefont {Mudrich}}, \bibinfo {author} {\bibfnamefont
			{K.}~\bibnamefont {Dulitz}}, \ and\ \bibinfo {author} {\bibfnamefont
			{T.}~\bibnamefont {Momose}},\ }\bibfield  {title} {\enquote {\bibinfo {title}
			{{Production of rotationally cold methyl radicals in pulsed supersonic
					beams}},}\ }\href {\doibase 10.1063/1.5052017} {\bibfield  {journal}
		{\bibinfo  {journal} {Rev. Sci. Instrum.}\ }\textbf {\bibinfo {volume}
			{89}},\ \bibinfo {pages} {113103} (\bibinfo {year} {2018})}\BibitemShut
	{NoStop}%
	\bibitem [{\citenamefont {Grzesiak}\ \emph {et~al.}(2019)\citenamefont
		{Grzesiak}, \citenamefont {Momose}, \citenamefont {Stienkemeier},
		\citenamefont {Mudrich},\ and\ \citenamefont {Dulitz}}]{Grzesiak2019}%
	\BibitemOpen
	\bibfield  {author} {\bibinfo {author} {\bibfnamefont {J.}~\bibnamefont
			{Grzesiak}}, \bibinfo {author} {\bibfnamefont {T.}~\bibnamefont {Momose}},
		\bibinfo {author} {\bibfnamefont {F.}~\bibnamefont {Stienkemeier}}, \bibinfo
		{author} {\bibfnamefont {M.}~\bibnamefont {Mudrich}}, \ and\ \bibinfo
		{author} {\bibfnamefont {K.}~\bibnamefont {Dulitz}},\ }\bibfield  {title}
	{\enquote {\bibinfo {title} {{Penning collisions between supersonically
					expanded metastable He atoms and laser-cooled Li atoms}},}\ }\href {\doibase
		10.1063/1.5063709} {\bibfield  {journal} {\bibinfo  {journal} {J. Chem.
				Phys.}\ }\textbf {\bibinfo {volume} {150}},\ \bibinfo {pages} {034201}
		(\bibinfo {year} {2019})}\BibitemShut {NoStop}%
	\bibitem [{\citenamefont {Hotop}(1996)}]{Hotop1996}%
	\BibitemOpen
	\bibfield  {author} {\bibinfo {author} {\bibfnamefont {H.}~\bibnamefont
			{Hotop}},\ }\bibfield  {title} {\enquote {\bibinfo {title} {{11. Detection of
					metastable atoms and molecules}},}\ }in\ \href {\doibase
		10.1016/S0076-695X(08)60793-0} {\emph {\bibinfo {booktitle} {Atomic,
				molecular, and optical physics: atoms and molecules}}},\ \bibinfo {series}
	{Experimental methods in the physical sciences}, Vol.\ \bibinfo {volume} {29,
		Part B},\ \bibinfo {editor} {edited by\ \bibinfo {editor} {\bibfnamefont
			{F.}~\bibnamefont {Dunning}}\ and\ \bibinfo {editor} {\bibfnamefont {R.~G.}\
			\bibnamefont {Hulet}}}\ (\bibinfo  {publisher} {Academic Press},\ \bibinfo
	{year} {1996})\ pp.\ \bibinfo {pages} {191 -- 215}\BibitemShut {NoStop}%
	\bibitem [{Note1()}]{Note1}%
	\BibitemOpen
	\bibinfo {note} {In this estimate, frequency drifts caused by thermal
		fluctuations inside the wavelength meter itself are disregarded.}\BibitemShut
	{Stop}%
	\bibitem [{\citenamefont {Sansonetti}\ \emph {et~al.}(1995)\citenamefont
		{Sansonetti}, \citenamefont {Richou}, \citenamefont {Engleman},\ and\
		\citenamefont {Radziemski}}]{Sansonetti1995}%
	\BibitemOpen
	\bibfield  {author} {\bibinfo {author} {\bibfnamefont {C.~J.}\ \bibnamefont
			{Sansonetti}}, \bibinfo {author} {\bibfnamefont {B.}~\bibnamefont {Richou}},
		\bibinfo {author} {\bibfnamefont {R.}~\bibnamefont {Engleman}}, \ and\
		\bibinfo {author} {\bibfnamefont {L.~J.}\ \bibnamefont {Radziemski}},\
	}\bibfield  {title} {\enquote {\bibinfo {title} {{Measurements of the
					resonance lines of $^{6}\mathrm{Li}$ and $^{7}\mathrm{Li}$ by Doppler-free
					frequency-modulation spectroscopy}},}\ }\href {\doibase
		10.1103/PhysRevA.52.2682} {\bibfield  {journal} {\bibinfo  {journal} {Phys.
				Rev. A}\ }\textbf {\bibinfo {volume} {52}},\ \bibinfo {pages} {2682}
		(\bibinfo {year} {1995})}\BibitemShut {NoStop}%
	\bibitem [{\citenamefont {Metcalf}\ and\ \citenamefont
		{Straten}(1999)}]{Metcalf1999}%
	\BibitemOpen
	\bibfield  {author} {\bibinfo {author} {\bibfnamefont {H.~J.}\ \bibnamefont
			{Metcalf}}\ and\ \bibinfo {author} {\bibfnamefont {P.~v.~d.}\ \bibnamefont
			{Straten}},\ }\href@noop {} {\emph {\bibinfo {title} {{Laser cooling and
					trapping}}}}\ (\bibinfo  {publisher} {Springer},\ \bibinfo {address} {New
		York},\ \bibinfo {year} {1999})\BibitemShut {NoStop}%
	\bibitem [{\citenamefont {Steck}(2001)}]{Steck2001}%
	\BibitemOpen
	\bibfield  {author} {\bibinfo {author} {\bibfnamefont {D.~A.}\ \bibnamefont
			{Steck}},\ }\emph {\bibinfo {title} {{Quantum chaos, transport, and
				decoherence in atom optics}}},\ \href@noop {} {Ph.D. thesis},\ \bibinfo
	{school} {The University of Texas at Austin} (\bibinfo {year}
	{2001})\BibitemShut {NoStop}%
	\bibitem [{\citenamefont {Budker}\ \emph {et~al.}(2004)\citenamefont {Budker},
		\citenamefont {Kimball}, \citenamefont {Kimball},\ and\ \citenamefont
		{DeMille}}]{Budker2004}%
	\BibitemOpen
	\bibfield  {author} {\bibinfo {author} {\bibfnamefont {D.}~\bibnamefont
			{Budker}}, \bibinfo {author} {\bibfnamefont {D.}~\bibnamefont {Kimball}},
		\bibinfo {author} {\bibfnamefont {D.}~\bibnamefont {Kimball}}, \ and\
		\bibinfo {author} {\bibfnamefont {D.}~\bibnamefont {DeMille}},\ }\href@noop
	{} {\emph {\bibinfo {title} {{Atomic physics: an exploration through problems
					and solutions}}}}\ (\bibinfo  {publisher} {Oxford University Press},\
	\bibinfo {address} {Oxford},\ \bibinfo {year} {2004})\BibitemShut {NoStop}%
	\bibitem [{\citenamefont {Kramida}\ \emph {et~al.}(2018)\citenamefont
		{Kramida}, \citenamefont {Ralchenko}, \citenamefont {Reader},\ and\
		\citenamefont {{NIST ASD Team}}}]{NIST_ASD}%
	\BibitemOpen
	\bibfield  {author} {\bibinfo {author} {\bibfnamefont {A.}~\bibnamefont
			{Kramida}}, \bibinfo {author} {\bibfnamefont {Y.}~\bibnamefont {Ralchenko}},
		\bibinfo {author} {\bibfnamefont {J.}~\bibnamefont {Reader}}, \ and\ \bibinfo
		{author} {\bibnamefont {{NIST ASD Team}}},\ }\href@noop {} {\enquote
		{\bibinfo {title} {{{NIST atomic spectra database} (Version 5.6)}},}\ }
	(\bibinfo {year} {2018}),\ \bibinfo {note}
	{\url{http://physics.nist.gov/asd}}\BibitemShut {NoStop}%
	\bibitem [{\citenamefont {Hollenstein}(2003)}]{Hollenstein2003}%
	\BibitemOpen
	\bibfield  {author} {\bibinfo {author} {\bibfnamefont {U.}~\bibnamefont
			{Hollenstein}},\ }\emph {\bibinfo {title} {{Erzeugung und spektroskopische
				Anwendungen von schmalbandiger, koh\"arenter, vakuum-ultravioletter
				Strahlung}}},\ \href@noop {} {Ph.D. thesis},\ \bibinfo  {school} {ETH Zurich}
	(\bibinfo {year} {2003})\BibitemShut {NoStop}%
	\bibitem [{\citenamefont {Phelps}(1955)}]{Phelps1955}%
	\BibitemOpen
	\bibfield  {author} {\bibinfo {author} {\bibfnamefont {A.~V.}\ \bibnamefont
			{Phelps}},\ }\bibfield  {title} {\enquote {\bibinfo {title} {{Absorption
					studies of helium metastable atoms and molecules}},}\ }\href {\doibase
		10.1103/PhysRev.99.1307} {\bibfield  {journal} {\bibinfo  {journal} {Phys.
				Rev.}\ }\textbf {\bibinfo {volume} {99}},\ \bibinfo {pages} {1307} (\bibinfo
		{year} {1955})}\BibitemShut {NoStop}%
\end{thebibliography}

%

\end{document}